\documentclass[acmsmall,screen,nonacm]{acmart} 
\setcopyright{none}
\usepackage{xspace}
\usepackage{booktabs}
\usepackage{tabularx}
\usepackage{multirow}
\usepackage{enumitem}
\usepackage[most]{tcolorbox}
\usepackage{tikz}
\usetikzlibrary{positioning,calc}

\newcommand{\flow}{\textsc{FlowPlan}\xspace}
\newcommand{\cc}{Claude Code\xspace}
\newcommand{\haiku}{Claude Haiku~4.5\xspace}
\newcommand{\tl}[1]{\texttt{#1}}
\newcommand{\xx}{$\times$}{}

\newcommand{\sig}{\textsuperscript{$\dagger$}}
\newcommand{\ns}{\textsuperscript{$\circ$}}

\newtcolorbox{findingbox}{
  enhanced,
  colback=blue!4!white,
  colframe=black,
  boxrule=0.6pt,
  arc=1.5pt,
  left=4pt, right=4pt, top=3pt, bottom=3pt,
  before skip=6pt, after skip=6pt
}

\begin{document}

\title{FlowPlan: Reducing the Token Cost of Coding Agents with Abstraction-Guided Planning and Verification}

\author{Melika Sepidband}
\email{melikasp@yorku.ca}
\author{Jiho Shin}
\email{shinjiho@yorku.ca}
\author{Song Wang}
\email{wangsong@yorku.ca}
\author{Hadi Hemmati}
\email{hemmati@yorku.ca}
\affiliation{%
  \institution{York University}
  \city{Toronto}
  \country{Canada}
}

\renewcommand{\shortauthors}{Sepidband et al.}

\begin{abstract}
LLM-based coding agents such as \cc can resolve real-world repository issues, but their iterative interaction loop can be expensive. For example, a single repair may consume millions of tokens as an increasingly long trajectory is replayed at every model call. Existing approaches reduce this cost primarily by limiting, compressing, or selectively avoiding information and actions within the repair process. We investigate a complementary question: Can an agent reason about a repair at an abstract program level, rather than repeatedly operating over raw source code, to reduce both the exploration needed to identify a repair and the repeated checking needed to verify it afterward?

We present \flow, an abstraction-guided repair approach that structures localization, planning, and verification around per-function control-flow or data-flow graphs. The agent first localizes the candidate fault location on the program abstraction and expresses the intended repair as explicit graph deltas before modifying the concrete source. After the edit, \flow reconstructs the same abstraction and compares the observed structural change with the planned one. Then, a verification checks the complete repair before termination. This process makes repair intent explicit before editing and provides a machine-checkable criterion for whether the intended structural change has been made.

Across all six benchmark--model settings, \flow does not significantly change resolution rate relative to the vanilla agent while reducing token consumption in every setting, significantly in five of six settings. With \haiku, it uses only 0.46--0.47\xx{} the baseline tokens on SWE-bench Pro, 0.32--0.35\xx{} on SWE-bench Verified, and 0.68--0.74\xx{} on SecBench; with Sonnet~5, it uses 0.72--0.76\xx, 0.73--0.74\xx, and 0.92--0.94\xx, respectively. The smaller reduction on SecBench reflects its substantially shorter baseline trajectories, which leave less accumulated context and post-edit work to eliminate. Overall, \flow achieves the largest token reduction among the compared methods in three of the six settings and ties the strongest compared method in a fourth. Trajectory analysis shows that \flow saves tokens by replacing much of the costly post-edit re-reading and testing with explicit verification and reducing per-call context. The main remaining challenges are precise fault localization and complete repair planning.

\end{abstract}

\ccsdesc[500]{Software and its engineering~Software maintenance tools}
\ccsdesc[300]{Software and its engineering~Software verification and validation}

\keywords{Coding agents, token efficiency, control-flow graph, data-flow graph, program repair}

\maketitle

\section{Introduction}
\label{sec:intro}

LLM-based coding agents have emerged as a practical approach to repository-level software engineering tasks. Given an issue, agents such as \cc~\cite{anthropic2026claudecode}, SWE-agent~\cite{yang2024swe}, and OpenHands~\cite{wang2025openhands} search repositories, inspect source files, edit code, and execute tests to produce a patch. This capability, however, can require substantial inference cost. In our baseline runs, with \haiku on SWE-bench Pro, the vanilla \cc consumes an average of 3.7 million tokens per issue, 98\% of which are cache reads. This cost comes directly from the agentic loop: each model call replays the conversation accumulated so far, and every tool observation enlarges it. Repair cost therefore has two drivers: the number of model calls and the context each call replays.

Recent work reduces this cost in several directions. AgentDiet prunes accumulated trajectories using a second model~\cite{agentdiet2026}. EET retrieves prior experience and encourages earlier termination~\cite{eet2026}. SHERLOC performs a dedicated localization stage before repair~\cite{sherloc2026}. \emph{To Run or Not to Run} removes test execution~\cite{noexec2026}. SeeRepo replaces part of textual repository exploration with a visual repository graph~\cite{seerepo2026}. These methods target different sources of overhead, but generally reduce or compress context or eliminate actions from the source-level repair process, leaving the agent to repeatedly inspect source code to determine where and how to edit and whether the repair has succeeded.

In this work, we investigate a different direction grounded in a fundamental software-engineering principle: \emph{program abstraction}. Software engineering uses abstractions to expose task-relevant program properties while suppressing lower-level details that need not be considered at every reasoning step. Control-flow graphs (CFGs), for example, expose how execution moves between program blocks, while data-flow graphs (DFGs) expose how values propagate from definitions to uses. We ask: \textit{Can using program abstraction before editing the source code reduce a coding agent's token consumption?} A coding agent reasons over source code and the context accumulated through its tool interactions. Even after locating the relevant code, it may continue reading files, inspecting diffs, and running tests while deciding how to repair the issue and when the repair is complete.

We introduce \flow, an abstraction-guided repair approach that uses a per-function CFG or DFG as the working level for localization, planning, and post-edit verification. Before editing, the agent anchors the candidate fault location to an element of the abstraction and expresses the intended repair as explicit \emph{graph deltas}, such as adding a guard, changing a branch condition, or introducing a definition--use relation. The agent then performs the source edit using its editing tools. Afterward, \flow reconstructs the same abstraction from the modified program and compares the observed structural change with the planned one. The abstraction therefore serves not merely as a compact view of code, but as the level at which the agent expresses its repair intent before editing and verifies afterward whether the resulting source change matches that intent.

\begin{figure*}[t!]
    \centering
    \includegraphics[width=0.8\textwidth]{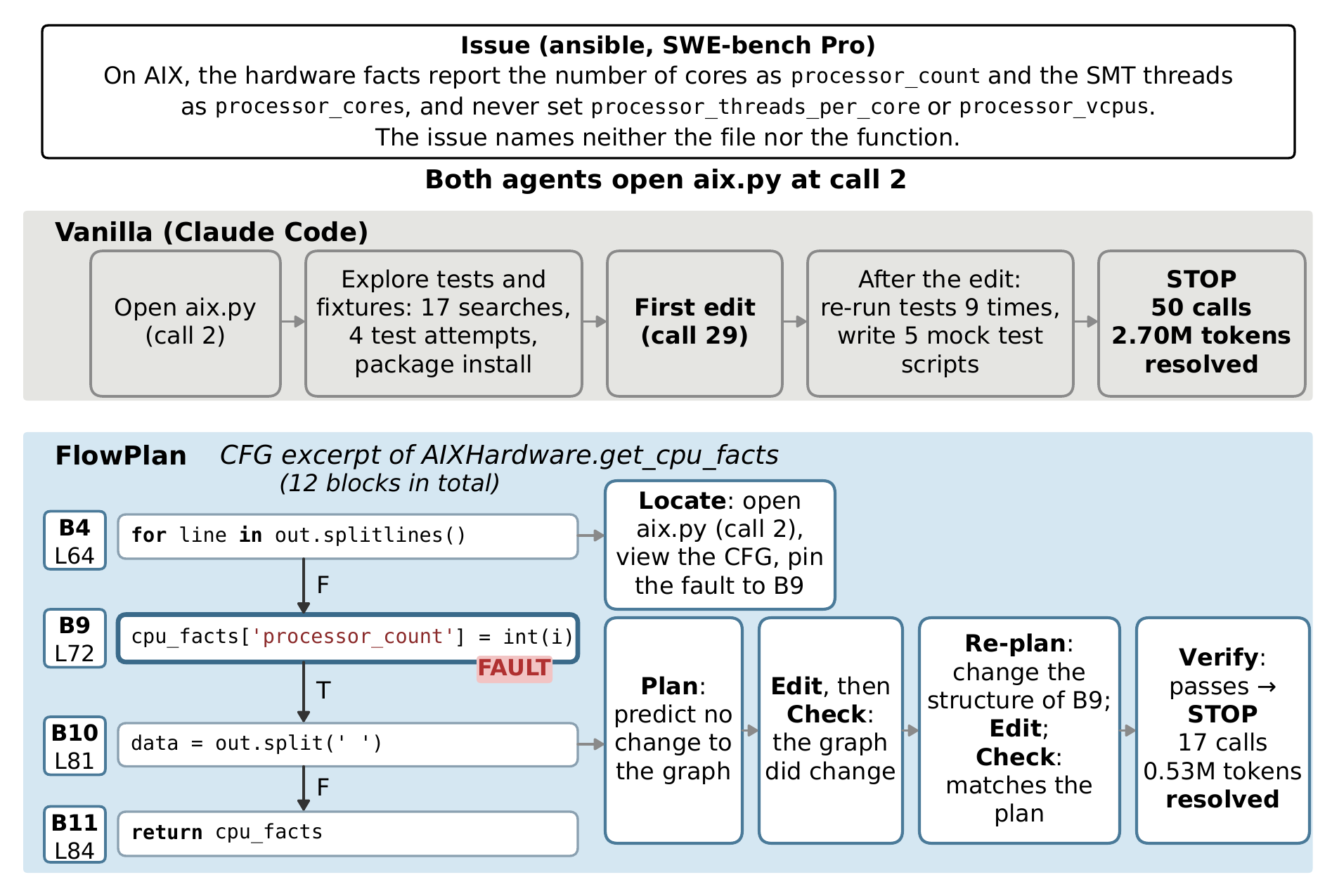}
    \caption{
    Motivating example from SWE-bench Pro (\tl{ansible-a6e671db}). Both agents identify the correct file at call~2, but their subsequent trajectories differ. The baseline continues repository exploration and repeated post-edit checking. \flow instead focuses the repair on the
    control-flow graph (CFG) of the relevant function. \texttt{B4}--\texttt{B11} denote basic blocks, \texttt{L64}--\texttt{L84} denote line number, and \texttt{T}/\texttt{F} denote true/false control-flow branches.
    It anchors the candidate fault location to a block, plans the intended CFG change, edits the code, and stops once
    the updated CFG matches that plan. Both agents resolve the issue, but \flow reduces the trajectory from 50 to 17 model calls and from 2.70M to 0.53M tokens.}
    \label{fig:motivation}
\end{figure*}

Figure~\ref{fig:motivation} illustrates this distinction on an issue from SWE-bench Pro. Both the vanilla \cc and \flow identify the correct file at their second model call, so the difference is not initial file localization. The baseline then spends 27 additional calls exploring related tests and fixtures before making its first edit at call~29, followed by another 21 calls repeatedly testing and checking the patch. It ultimately resolves the issue in 50 calls and 2.70M tokens. In contrast, \flow focuses the repair on the CFG of the relevant function: it identifies the candidate faulty block, states the intended structural change, edits the source, and verifies that the resulting CFG matches that plan. It resolves the same issue in 17 calls and 0.53M tokens. This example reflects a broader pattern: on SWE-bench Pro with \haiku, approximately two-thirds of the baseline's tokens are consumed after the first substantive edit through repeated re-reading, diff inspection, and testing.

\flow organizes this process in four phases: \emph{locate, plan, edit-and-check, and verify}. A final verification checks the complete repair before termination. Note that \flow does not replace or restrict the agent's ordinary file, search, editing, or shell tools. The agent can still fall back to them when additional context is needed.



We evaluate \flow with \cc and two models, \haiku and Claude Sonnet~5, on three benchmarks: SWE-bench Pro~\cite{deng2025swe} and SWE-bench Verified~\cite{openai2024sweverified} for Python program repair, and SecBench~\cite{lee2026sec} for C/C++ security patching. We evaluate 158, 218, and 169 instances, respectively, sized for a 95\% confidence level and a 5\% margin of error. To place \flow among recent token-reduction approaches, we survey 171 papers from twelve 2026 software-engineering and machine-learning venues and reproduce five methods using the same instances, models, \cc build, and accounting scheme while retaining their implementations, prompts, and thresholds.

Across the six benchmark--model settings, \flow does not significantly change resolution relative to the vanilla agent while reducing token consumption in every setting, significantly so in five of six. With \haiku, it uses 0.46--0.47\xx{} the baseline tokens on SWE-bench Pro, 0.32--0.35\xx{} on SWE-bench Verified, and 0.68--0.74\xx{} on SecBench; with Sonnet~5, it uses 0.72--0.76\xx, 0.73--0.74\xx, and 0.92--0.94\xx, respectively. The smaller reduction on SecBench reflects its substantially shorter baseline trajectories, which leave less accumulated context and post-edit work to eliminate. \flow achieves the largest token reduction among the compared methods in three of the six settings and ties the strongest compared method in a fourth. Trajectory analysis shows that \flow saves tokens by reducing the context replayed per call and by replacing much of the costly post-edit re-reading and testing with explicit verification. The main remaining challenges are precise fault localization on the graph and complete repair planning.

\smallskip
\noindent This work makes the following contributions:
\begin{itemize}[itemsep=2pt,topsep=2pt]
    \item \textbf{An abstraction-guided repair approach for coding agents.}
    \flow uses per-function control- and data-flow graphs as the level at which the agent localizes the candidate fault, expresses the intended repair as graph deltas, and verifies after editing whether the resulting source change produces the planned structural transformation.

    \item \textbf{A repair policy with an explicit completion criterion.}
    A four-phase \emph{locate--plan--edit-and-check--verify} policy structures how the agent uses the abstraction throughout the repair and uses graph-delta verification to determine when the intended change has been made, reducing otherwise open-ended post-edit checking.

    \item \textbf{A controlled comparison of token-reduction methods.}
    We survey 171 papers from twelve 2026 venues and reproduce five applicable token-reduction methods alongside \flow on three benchmarks and two models under the same agent environment and accounting scheme, measuring task success and token consumption broken down into input, output, cache-read, and cache-write tokens.

    \item \textbf{An analysis of protocol correctness and the sources of token savings.}
    We evaluate each stage of the \flow protocol against the reference repair and decompose token consumption by trajectory phase, model calls, and context per call, identifying where the savings arise and when they diminish.
\end{itemize}

\section{Background and Related Work}
\label{sec:background}

\subsection{Agentic Software Engineering}
\label{sec:background:agents}

LLM-based coding agents solve repository-level software engineering tasks by interleaving model inference with tool use. Systems such as SWE-agent~\cite{yang2024swe}, OpenHands~\cite{wang2025openhands}, AutoCodeRover~\cite{zhang2024autocoderover}, and \cc~\cite{anthropic2026claudecode} allow the model to search and read repositories, edit files, and execute shell commands. After each tool call, the resulting observation is added to the conversation, and the model chooses the next action until it decides that the task is complete. Benchmarks such as SWE-bench~\cite{jimenez2023swe}, SWE-bench Verified~\cite{openai2024sweverified}, SWE-bench Pro~\cite{deng2025swe}, and SecBench~\cite{lee2026sec} evaluate these agents on real repository issues and vulnerabilities by testing their final patches.

This iterative interaction pattern also drives the agent's token consumption. Because each model call receives the conversation accumulated so far, token usage grows with both the number of calls and the amount of context replayed at each call. Prompt caching can reduce the monetary cost of repeatedly processed prefixes, but it does not reduce the number of tokens processed.

\subsection{Code Representations and Program Abstractions}
\label{sec:background:rep}

A code representation makes program structure that is implicit in source text explicit, with different abstractions preserving different relations. Abstract syntax trees (ASTs) represent the syntactic structure of code, including declarations, expressions, and their nesting relationships~\cite{mou2016convolutional}. Call graphs capture potential calling relationships between functions, while dependency graphs capture coarser-grained relations such as imports and dependencies among files or modules. Repository-level representations often combine several such relations to support code navigation, retrieval, and localization~\cite{ouyang2025repograph,luo2026closing,chen2025locagent,seddik2026arise,codegraph,seerepo2026, sepidband2026should}. In contrast, control-flow and data-flow graphs characterize behavior within a program unit: a control-flow graph (CFG) represents the possible execution paths between basic blocks~\cite{allen1970control}, while a data-flow graph (DFG) represents how values propagate from definitions to uses; program dependence graphs combine control and data dependences~\cite{ferrante1987program}. These are longstanding software-engineering abstractions used in program analysis and compilation. They have also been incorporated into modern code-analysis and learning systems: code property graphs combine syntactic, control-flow, and data-flow information for vulnerability analysis~\cite{yamaguchi2014modeling}, GraphCoder retrieves completion context using statement-level control-flow and dependence relations~\cite{liu2024graphcoder}, and learned code models incorporate flow information as input~\cite{allamanis2017learning,guo2020graphcodebert,zhou2019devign,cummins2021programl}.

\subsection{\cc, Skills, and MCP Tools}
\label{sec:background:mcp}

\cc is Anthropic's command-line coding agent. Its built-in interface provides tools such as \tl{Read}, \tl{Edit}, \tl{Write}, \tl{Bash}, \tl{Grep}, and \tl{Glob}. In headless execution, it also reports input, output, cache-read, and cache-write tokens.

We use two \cc extension mechanisms. First, the Model Context Protocol (MCP)~\cite{anthropic2024mcp} allows an external server to expose additional tools. Their names, descriptions, and input schemas become part of the model's prompt prefix, and their outputs are returned as ordinary tool observations. Second, a \emph{skill} packages instructions that are inserted into the conversation when invoked. \flow uses an MCP server for its flow-graph tools and a skill for its repair policy (Section~\ref{sec:approach}). Because MCP tool descriptions occupy the prompt prefix on every call, they contribute to token consumption even when a tool is not invoked.

\subsection{Token Reduction for Agentic Software Engineering}
\label{sec:background:rw}

\textit{\textbf{Scope and comparator selection.}}
To identify recent approaches for comparison, we survey papers published or accepted in 2026 at twelve software-engineering and machine-learning venues: ICSE, FSE, ASE, TSE, TOSEM, NeurIPS, ICLR, ICML, AAAI, EMNLP, ISSTA, and ACL. The search was completed on 9 September 2026. For the two journals, TSE and TOSEM, which publish continuously, the search covers the papers that appeared or were accepted in 2026 up to that date. Among 171 confirmed papers, 27 treat reducing the tokens, context, cost, or interaction length of a software-engineering agent as a central contribution, and 18 release official implementations. We exclude approaches whose primary contribution is a trained model or that require model internals unavailable through \cc, such as token log-probabilities.

Thus, we reproduce five approaches that intervene at distinct points of the repair trajectory: SHERLOC~\cite{sherloc2026}, AgentDiet~\cite{agentdiet2026}, SeeRepo~\cite{seerepo2026}, EET~\cite{eet2026}, and \emph{To Run or Not to Run}~\cite{noexec2026}, hereafter \emph{No-execution}. Table~\ref{tab:rwcompare} summarizes where each method intervenes and how it differs from \flow.

\begin{table}[t]
\centering
\caption{
Comparison of the reproduced token-reduction methods.
\emph{Stage} indicates where the method intervenes in the repair trajectory. \emph{Program view} is the representation exposed to the repair agent. \emph{Abstract plan} indicates whether the intended repair is explicitly expressed over that representation before editing. \emph{Aux.\ LLM} indicates additional model calls outside the main repair loop (EET uses them only offline to construct its experience base). \emph{Stop} indicates whether the method provides an explicit completion criterion.}
\vspace{-0.1in}
\label{tab:rwcompare}
\footnotesize
\setlength{\tabcolsep}{3pt}
\begin{tabularx}{\columnwidth}{
@{}l
>{\raggedright\arraybackslash}X
>{\raggedright\arraybackslash}X
ccc@{}
}
\toprule
Method & Stage & Program view & Abstract plan & Aux.\ LLM & Stop \\
\midrule

SHERLOC~\cite{sherloc2026}
    & before repair
    & file tree, imports
    & -- & \checkmark & -- \\

AgentDiet~\cite{agentdiet2026}
    & after each step
    & --
    & -- & \checkmark & -- \\

SeeRepo~\cite{seerepo2026}
    & exploration
    & repository graph
    & -- & -- & -- \\

EET~\cite{eet2026}
    & start, key steps
    & --
    & -- & offline & soft \\

No-execution~\cite{noexec2026}
    & testing
    & --
    & -- & -- & -- \\

\midrule

\flow (ours)
    & locate, plan, verify
    & per-function CFG/DFG
    & \checkmark & -- & \checkmark \\

\bottomrule
\end{tabularx}
\end{table}

SHERLOC~\cite{sherloc2026} (EMNLP 2026) reduces pre-repair exploration through a dedicated localization agent whose findings are passed to the repair agent; we reproduce its two-stage workflow in \cc and include both sessions in the token accounting. AgentDiet~\cite{agentdiet2026} (FSE 2026) periodically uses a second LLM to rewrite and compress the accumulated trajectory; we reproduce its trigger, acceptance rule, and reflection prompt, and include the auxiliary reflection calls in the reported tokens. SeeRepo~\cite{seerepo2026} (ASE 2026) replaces part of textual repository exploration with an AST-derived graph queried as images; we expose its graph builder and renderer through MCP, retain its prompting behavior, and evaluate it only on Python because the released implementation supports that language. EET~\cite{eet2026} (ACL 2026 Findings) retrieves prior repair experience and uses confidence estimates after key events to encourage earlier termination; we retain its retrieval and stopping mechanisms. To avoid leakage, its experience base contains trajectories only from instances outside our evaluation sample: for SWE-bench Pro and SecBench, we construct it from disjoint baseline instances, while for SWE-bench Verified, we use the authors' released non-overlapping experience base. Finally, \emph{To Run or Not to Run}~\cite{noexec2026} (ISSTA 2026) reduces agent interaction by prohibiting program execution during repair; we reproduce this condition by adding the released no-execution instruction to the task prompt and withholding project dependencies. For SecBench's C/C++ projects, we adapt only the language-specific components necessary to preserve each method's intervention.

\textit{\textbf{Other related directions.}}
The reproduced methods represent broader approaches to reducing agent cost. \emph{Context-management} methods compress or selectively retain trajectory history, sometimes using learned policies~\cite{liu2026contextastool,lian2026sweagile,sun2026contextfolding,kang2026acon}. General-purpose prompt-compression techniques instead reduce individual prompts~\cite{pan2024llmlingua,li2023compressing}. \emph{Localization} methods shorten repository exploration through dedicated search agents or learned retrieval policies~\cite{xu2026fusesearch,cho2026long,mamun2026blagent}, building on earlier work such as AutoCodeRover, OrcaLoca, and LocAgent~\cite{zhang2024autocoderover,yu2025orcaloca,chen2025locagent}. \emph{Repository representations} expose structural views of a code base to support navigation and reasoning~\cite{ouyang2025repograph,chen2025locagent,seddik2026arise,luo2026closing}, while program graphs have also been widely studied as learned representations of source code~\cite{allamanis2017learning,guo2020graphcodebert}.
Finally, orthogonal work studies resource allocation rather than reducing context or actions within an unconstrained repair trajectory: fixed or adaptive turn budgets constrain how much interaction a task receives~\cite{gao2026morewithless}, budget-aware agents learn to estimate whether sufficient resources remain~\cite{lin2026bagen}, and test-time scaling allocates computation selectively across candidate solutions~\cite{mao2026egss}.

\flow differs primarily in how the program representation is used. Existing repository representations are typically introduced to support code discovery or navigation. After the relevant code is found, repair reasoning returns to source text. In \flow, the per-function CFG or DFG serves as the abstraction on which the agent localizes the candidate fault, formulates the intended structural change, and later verifies the effect of the concrete source edit. The representation therefore supports both pre-edit reasoning and post-edit verification, rather than only retrieval. Among the reproduced methods, EET also targets termination, but it relies on the agent's self-reported confidence, whereas \flow checks whether the observed program change matches the previously stated repair intent.
\section{\flow}
\label{sec:approach}

\begin{figure*}[t]
\centering
\includegraphics[width=\textwidth]{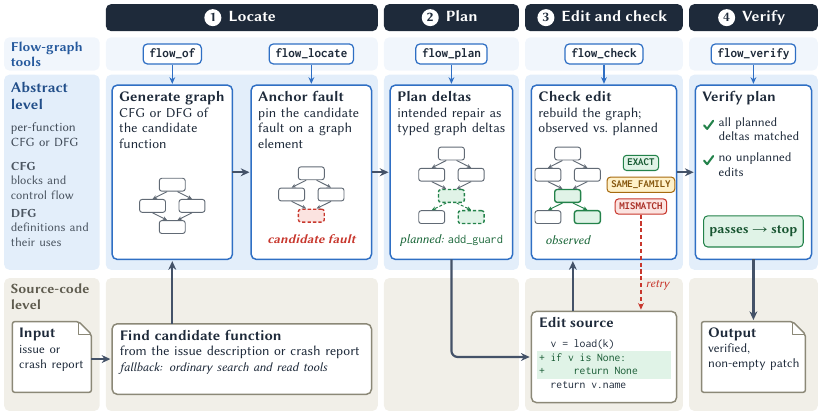}
\caption{
Overview of \flow. Columns show the four repair phases, while rows separate the flow-graph tools, abstract representation, and source code. The agent localizes the fault on a CFG or DFG, plans the repair as graph deltas, edits the source, and verifies that the observed graph changes match the plan.
}
\label{fig:overview}
\end{figure*}

Figure~\ref{fig:overview} summarizes \flow. Its central design is to perform key repair decisions at an abstract program level rather than directly over source text. Given a candidate function, a \emph{flow-graph engine} constructs a per-function CFG or DFG that exposes the structural relations relevant to the repair (Section~\ref{sec:approach:views}). The agent localizes the candidate fault and plans the intended repair on this abstraction before editing the concrete source. Five tools maintain the connection between the abstract repair plan and the source changes that follow (Section~\ref{sec:approach:tools}), while a four-phase policy organizes their use from localization through final verification (Section~\ref{sec:approach:protocol}).

The tools are attached to an otherwise vanilla \cc session through the Model Context Protocol (MCP)~\cite{anthropic2024mcp}. The agent retains its file-reading, search, editing, and shell tools throughout the repair. \flow therefore does not replace source-level interaction. Instead, the CFG or DFG provides an abstract working representation for localization, planning, and verification, while the agent can return to the source or execute commands whenever additional information is needed.

\subsection{Per-Function Program Abstraction}
\label{sec:approach:views}

\flow constructs program abstraction at the granularity of individual functions. Rather than encoding the entire repository as a graph, the flow-graph engine constructs an on-demand view of the function currently under investigation. This keeps the abstraction focused on the local structure needed for repair while omitting unrelated repository content, so the agent can reason about the candidate fault and intended change before modifying the source code.

For a control-flow graph (CFG), the view contains the function's basic blocks, their source locations, and the control-flow edges between them. Edges are annotated with their roles, such as true, false, loop, return, or raise. For a data-flow graph (DFG), the view records variable definitions and the uses reached by each definition, together with their source locations. For Python, both views are derived from the abstract syntax tree. For C and C++, they are extracted from a code-property graph constructed with Joern and restricted to functions relevant to the crash stack.

Each view also induces a canonical fingerprint. For a CFG, the fingerprint captures the normalized block and edge set. For a DFG, it captures the multiset of definition--use relations. \flow records the fingerprint of the original function and recomputes it after the source is edited. The difference between these fingerprints defines the observed graph delta: the structural change that the edit actually introduced. Because the repair plan is expressed using the same graph vocabulary, the observed delta can be compared directly with the planned graph delta specified by the agent.

\textit{\textbf{Symbol index.}} To construct a requested abstraction, \flow must first map the function reference supplied by the agent to a concrete source definition. Before each session, it therefore builds a repository-level symbol index from the instance's base commit using tree-sitter. Each entry stores its name, qualified name, kind, file, and line range in SQLite. The index is built once per repository commit, outside the agent session and without model calls. 

\subsection{Flow-Graph Tools}
\label{sec:approach:tools}

\begin{table}[t]
\centering
\caption{The five \flow tools exposed to the agent. A server provides either CFG or DFG views. The tool interface and repair policy are identical across representations.}
\label{tab:tools}
\small
\setlength{\tabcolsep}{3pt}
\begin{tabularx}{\columnwidth}{@{}l>{\raggedright\arraybackslash}X@{}}
\toprule
Tool & Role \\
\midrule
\tl{flow\_of} & Resolve a function through the symbol index and return its CFG or DFG with source locations. \\
\tl{flow\_locate} & Anchor the candidate fault to a validated graph element and suggest nearby valid candidates when the reference is invalid. \\
\tl{flow\_plan} & Record the intended repair as typed graph deltas and snapshot graph fingerprints from \tl{HEAD}. \\
\tl{flow\_check} & Recompute an edited function's graph and compare the change with its planned delta. \\
\tl{flow\_verify} & Verify the complete repair plan and detect modified functions that were not in the plan. \\
\bottomrule
\end{tabularx}
\end{table}

Table~\ref{tab:tools} summarizes the five tools exposed to the agent. Together, they span the repair from initial localization to final verification. They also link the fault the agent identifies, the change it intends to make, and the structural change its edit produces.

\textit{\textbf{Localization.}} \tl{flow\_of} takes a function name, optionally qualified by a file, resolves it through a repository-level symbol index, and returns the corresponding flow view. Because resolution is symbol-based, an agent that already knows a candidate function from the issue description or crash report can inspect its structure without first performing a textual repository search.

After inspecting the view, the agent calls \tl{flow\_locate} to identify the \emph{fault location}: the graph element that it believes must change to repair the issue. Depending on the selected abstraction, this site may be a basic block, an edge, a definition--use relation, or a source line. \tl{flow\_locate} validates the supplied reference against the generated graph.

This validation does not prevent the repair from proceeding when the agent localizes incorrectly. If the supplied reference is invalid, \tl{flow\_locate} records the attempt and returns the three nearest valid candidates so that the agent can correct the reference. After two unsuccessful attempts, the location is retained but marked as unverified. We therefore treat localization as an observable reasoning decision rather than a hard execution constraint: localization errors remain measurable without artificially terminating the repair.

\textit{\textbf{Planning.}} Once a fault location has been selected, \tl{flow\_plan} records the intended repair as one or more typed \emph{graph deltas}. Each delta identifies the target function, an anchor line, the fault location it addresses, a rationale, and the structural operation the agent expects the edit to induce.

For CFGs, the operation vocabulary includes \tl{add\_guard}, \tl{remove\_branch}, \tl{change\_condition}, and \tl{rewire\_edge}. For DFGs, operations describe additions, removals, or movements of definition--use relations. Two operations apply independently of the selected representation. \tl{no\_flow\_change} represents a repair that modifies the source without changing the chosen flow graph. Such a delta therefore includes a source anchor against which the edit can still be checked. \tl{add\_function} represents a repair that introduces a function absent from the original program.

When a delta is first recorded, \flow snapshots the fingerprint of its target function from the repository's \tl{HEAD} state, rather than from the current working tree. This distinction is important because the agent may edit the file before invoking \tl{flow\_plan}. Using the version-controlled state ensures that verification still compares the final program against the pre-repair structure.

Planning is cumulative rather than destructive. Subsequent calls to \tl{flow\_plan} merge new deltas by function and append the plan to history instead of replacing the previous plan. This preserves re-planning decisions and allows us to reconstruct how the agent's intent evolved during the trajectory.

\textit{\textbf{Verification.}} After an edit, \tl{flow\_check} reconstructs the graph from the current working tree and compares its fingerprint with the snapshot of the original function. This comparison yields the observed graph delta, which is then evaluated against the corresponding planned delta.

The comparison produces one of three graded outcomes. \textsc{exact} indicates that the planned structural operation is directly present in the observed graph delta. \textsc{same\_family} indicates that the edit produces the intended structural effect through a related operation. For example, an edge rewiring may produce the effect anticipated by a planned condition change. \textsc{mismatch} indicates that the observed graph change does not match the planned delta. This graded comparison avoids treating structurally equivalent implementations as failures solely because their low-level graph operations differ.

\tl{flow\_verify} applies the same comparison to the repair plan before termination. It additionally audits the working tree for modified functions that are absent from the plan, ensuring that verification covers the patch rather than only the most recently checked function. For an \tl{add\_function} delta, no original fingerprint exists at \tl{HEAD}. The verifier therefore extracts the new function from the current version of its anchor file and succeeds if a valid fingerprint can be constructed.

\subsection{Four-Phase Repair Policy}
\label{sec:approach:protocol}

The graph abstraction and tools do not by themselves determine when the agent should localize, commit to a plan, or decide that the repair is complete. \flow therefore operationalizes the abstraction-guided repair with a four-phase repair policy: \emph{locate, plan, edit-and-check, verify}. The policy is delivered as a \cc skill, \tl{/flowfix}, whose instructions are expanded into the initial user message. The skill is generated from a common source across benchmarks and graph views, stored outside the target repository, and verified by the evaluation harness to ensure that it is loaded in every run.
The four phases are:

\begin{enumerate}[itemsep=2pt,topsep=2pt]
    \item \textbf{Locate.} The agent first identifies the most likely faulty function using evidence already available in the issue description or crash report and calls \tl{flow\_of} to inspect its graph. If the view does not expose the fault, the agent may fall back to its standard search and file-reading tools, select another candidate function, and inspect its graph. Once a candidate is chosen, the agent anchors the candidate fault location using \tl{flow\_locate}.
    \item \textbf{Plan.} Before editing, the agent calls \tl{flow\_plan} and expresses the intended repair as one or more graph deltas. The plan must cover every function the agent expects to modify and associate each delta with the fault location it is intended to address. This makes the intended structural effect explicit at the abstract level before the source code is changed.
    \item \textbf{Edit and check.} The agent makes the smallest source edit intended to produce the planned delta and invokes \tl{flow\_check} on the modified function. The tool reconstructs the abstraction and checks whether the concrete edit produced the planned structural change. If the result is \textsc{mismatch}, the policy allows one additional edit-and-check attempt. If the second check still does not match the plan, the agent may revise the plan once before continuing.
    \item \textbf{Verify.} Before terminating, the agent invokes \tl{flow\_verify} over the complete repair plan and modified working tree. If verification fails, the policy permits one corrective edit followed by one final verification. A successful verification together with a non-empty patch is the policy's completion criterion.
\end{enumerate}

The policy is intentionally advisory rather than mechanically enforced. No runtime hook forces the agent to invoke a particular tool, disables its native tools, or prevents it from terminating. Instead, we record whether the agent follows each phase as part of the trajectory analysis. This preserves the behavior of the underlying coding agent while allowing the abstraction to guide its repair process without imposing a constrained execution environment.

The retry limits bound the additional interaction introduced by verification. A failed check or verification should trigger a bounded attempt to reconcile the implementation with the plan, not restart an open-ended repair loop. Allowing one edit and, when necessary, one re-planning step allows the agent to recover from a mismatch while limiting the interaction that verification itself can introduce.

\subsection{Why \flow Is Expected to Reduce Cost}
\label{sec:approach:policy}

\flow targets two costly parts of an agent trajectory. Before editing, reasoning over a focused per-function abstraction can reduce repeated repository exploration and the source context accumulated in the trajectory. After editing, comparing the reconstructed abstraction with the planned graph deltas provides an explicit completion criterion that can reduce repeated source inspection, diff checking, and testing. RQ3 evaluates whether these mechanisms explain the observed token savings.
\section{Experimental Setup}
\label{sec:setup}

\subsection{Research Questions}
\label{sec:setup:rqs}

We evaluate \flow through the following research questions:

\begin{itemize}[itemsep=2pt,topsep=2pt]
    \item \textbf{RQ1 (Effectiveness and cost).} How does \flow affect task success and token consumption relative to the vanilla coding agent?
    \item \textbf{RQ2 (Comparison with prior token-reduction methods).} How does \flow compare with existing token-reduction methods when they are evaluated on the same instances, models, agent environment, and accounting scheme?
    \item \textbf{RQ3 (Mechanism analysis).} How reliably does each stage of the \flow protocol perform its intended role, and where in the agent trajectory do the token savings arise?
\end{itemize}

\subsection{Agent and Models}
\label{sec:setup:agent}

We use \cc version 2.1.220 with two models that differ in capability and price: \haiku and Claude Sonnet~5, both configured with low reasoning effort. Each benchmark instance is executed in a fresh headless \cc session for every method--model configuration, preventing conversational state from carrying across runs. Each method is run once per instance.

Our baseline is the vanilla \cc with its default tool interface. Reproduced methods retain this interface unless their design introduces additional tools or stages (Section~\ref{sec:background:rw}). For \flow, we attach the flow-graph server through MCP and prepend the \tl{/flowfix} skill that implements the four-phase policy in Section~\ref{sec:approach:protocol}. CFG and DFG are evaluated as separate \flow configurations. 

\subsection{Benchmarks}
\label{sec:setup:benchmarks}

We evaluate on three benchmarks spanning repository-level Python repair and C/C++ security patching. For each benchmark, we draw a sample sized for a 95\% confidence level and a 5\% margin of error~\cite{cochran_sampling}, using a fixed random seed selected before observing any experimental outcome. 

\textit{\textbf{SWE-bench Pro}}~\cite{deng2025swe} contains 266 public Python instances from \emph{ansible}, \emph{openlibrary}, and \emph{qutebrowser} repositories. Our 158-instance sample contains 57, 55, and 46 instances, respectively.

\textit{\textbf{SWE-bench Verified}}~\cite{openai2024sweverified} is the human-validated subset of 500 SWE-bench~\cite{jimenez2023swe} instances. Our 218-instance sample spans eleven repositories, and the most frequent are \emph{django} (95 instances), \emph{sympy} (33), and \emph{scikit-learn} and \emph{sphinx} (17 each).

\textit{\textbf{SecBench}}~\cite{lee2026sec} contains vulnerabilities in C/C++ projects accompanied by OSS-Fuzz sanitizer reports and proof-of-concept inputs. Given the repository and failure information, the agent must modify the program so that the proof-of-concept input no longer triggers the sanitizer while preserving behavior. We evaluate 169 of the 300 instances in the benchmark's evaluation split.

Every evaluated method receives the same task description for a given instance. For SWE-bench Pro, we pass the \tl{problem\_statement}, \tl{requirements}, and \tl{interface} fields. For SWE-bench Verified, we pass the \tl{problem\_statement}. For SecBench, we pass the original bug report and the sanitizer report. No arm receives the reference patch or the names of the fail-to-pass tests.

We evaluate every generated patch using the benchmark's official evaluation harness. For SWE-bench Pro and Verified, an instance is \emph{resolved} when all fail-to-pass tests pass while the pass-to-pass tests continue to pass. For SecBench, an instance is resolved when the benchmark harness no longer observes the target sanitizer failure on the proof-of-concept input.

Each session is capped at 80 turns, \$2, and 25 minutes on SWE-bench, and 40 turns, \$1.50, and 25 minutes on SecBench, following the benchmark evaluation settings~\cite{deng2025swe, lee2026sec}. These limits are well above typical baseline usage and therefore act mainly as safeguards against unusually long runs. The same limits apply to all methods, and runs that reach a cap are graded on the patch produced at that point.

\subsection{Metrics}
\label{sec:setup:metrics}

For each session, we measure \textbf{task success} and \textbf{token consumption}. Task success indicates whether the generated patch resolves the benchmark instance under its official evaluation harness. Token consumption is the sum of four provider-reported components across all model calls required by a method: \emph{input} tokens, \emph{cache-write} tokens, \emph{cache-read} tokens, and \emph{output} tokens. Input tokens are uncached prompt tokens, cache-write tokens are prompt tokens newly stored in the cache, cache-read tokens are prompt tokens retrieved from the cache, and output tokens are generated by the model. Our accounting includes auxiliary calls, such as separate localization or reflection stages, when required by a method.

We report the mean and median total tokens per instance and the mean of each token component. For comparison across methods, we normalize each method's mean total tokens by the mean of the vanilla agent on the same instances; for example, a ratio of 0.45\xx means that the method uses 45\% as many tokens as the baseline.

All comparisons are paired by benchmark instance and two-sided. We compare task success using McNemar's exact test. Because our primary token metric is the mean, we compare token consumption using a paired permutation test on the mean per-instance difference. Under the null hypothesis, method labels are exchangeable within each instance; we therefore randomly swap each paired observation and recompute the mean difference over 200{,}000 permutations. The $p$-value is the proportion of permuted differences at least as extreme in absolute value as the observed difference~\cite{smucker2007comparison}. Because we compare every method in every benchmark--model setting, we apply the Holm--Bonferroni correction across all 40 token comparisons.
\section{Results}
\label{sec:results}

\providecommand{\sig}{\textsuperscript{$\dagger$}}
\providecommand{\ns}{\textsuperscript{$\circ$}}

\begin{table*}[t]
\centering
\caption{
Task success and token consumption of \flow and reproduced token-reduction methods using \cc at low reasoning effort. Baseline is vanilla \cc. Resolve rate is the percentage of resolved instances. Token columns report mean tokens per instance in thousands; Total median and Total mean aggregate the four token categories. Ratio is the method's Total mean divided by the baseline Total mean on matched instances. Bold marks the lowest ratio per benchmark-model setting. Ratios are compared using paired two-sided permutation tests with Holm correction over 40 comparisons ($p<0.05$); \ns denotes non-significance, and \sig{} a significant resolution difference under McNemar's exact test. SHERLOC includes localization and AgentDiet reflection calls. EET uses disjoint baseline trajectories or the released experience base for SWE-bench Verified. SeeRepo is Python-only and therefore omitted on SecBench.
}
\label{tab:main}
\scriptsize
\setlength{\tabcolsep}{1.5pt}
\begin{tabular}{@{}l r rrrrrrr@{\hspace{5pt}}r rrrrrrr@{}}
\toprule
& \multicolumn{8}{c}{\haiku} & \multicolumn{8}{c}{Claude Sonnet~5} \\
\cmidrule(lr){2-9}\cmidrule(l){10-17}
& & \multicolumn{7}{c}{Tokens per instance (thousands)} & & \multicolumn{7}{c}{Tokens per instance (thousands)} \\
\cmidrule(lr){3-9}\cmidrule(l){11-17}
Method & Resolve & Input & Output & Cache & Cache & Total & Total & Ratio & Resolve & Input & Output & Cache & Cache & Total & Total & Ratio \\
& rate (\%) & & & read & write & median & mean &  & rate (\%) & & & read & write & median & mean &  \\
\midrule
\multicolumn{17}{@{}l}{\emph{SWE-bench Pro ($n=158$)}} \\
Baseline & 43.0 & 0.3 & 16.8 & 3,634 & 63.6 & 3,528 & 3,715 & -- & 52.5 & $<$0.1 & 6.5 & 1,118 & 27.8 & 850 & 1,152 & -- \\
\addlinespace[2pt]
No-execution & 48.1 & 0.2 & 14.1 & 2,955 & 57.6 & 2,803 & 3,027 & 0.81\xx & 50.6 & $<$0.1 & 5.7 & 874 & 25.5 & 613 & 905 & 0.79\xx \\
EET & 46.2 & 0.3 & 17.0 & 3,249 & 91.2 & 3,084 & 3,357 & 0.90\xx & 63.3\sig & $<$0.1 & 6.3 & 1,112 & 27.7 & 843 & 1,146 & 0.99\xx\ns \\
SHERLOC & 48.1 & 0.3 & 18.0 & 3,317 & 137.4 & 3,174 & 3,472 & 0.93\xx\ns & 58.2 & $<$0.1 & 7.3 & 964 & 62.5 & 807 & 1,034 & 0.90\xx\ns \\
AgentDiet & 40.5 & 131.3 & 109.2 & 2,639 & 180.3 & 3,011 & 3,060 & 0.82\xx & 51.9 & 37.6 & 11.7 & 1,192 & 55.5 & 1,067 & 1,297 & 1.13\xx\ns \\
SeeRepo & 44.9 & 0.3 & 18.5 & 4,033 & 72.0 & 3,911 & 4,123 & 1.11\xx & 56.3 & $<$0.1 & 6.0 & 1,074 & 30.2 & 874 & 1,111 & 0.96\xx\ns \\
\midrule
\flow CFG & 46.8 & 0.3 & 14.6 & 1,660 & 80.6 & 1,412 & 1,755 & 0.47\xx & 56.3 & $<$0.1 & 6.9 & 798 & 30.3 & 616 & 835 & \textbf{0.72\xx} \\
\flow DFG & 46.2 & 0.3 & 14.4 & 1,616 & 82.8 & 1,351 & 1,713 & \textbf{0.46\xx} & 57.0 & $<$0.1 & 7.3 & 838 & 30.6 & 657 & 876 & 0.76\xx \\
\midrule
\multicolumn{17}{@{}l}{\emph{SecBench ($n=169$)}} \\
Baseline & 29.0 & 0.1 & 4.0 & 734 & 28.9 & 501 & 767 & -- & 37.9 & $<$0.1 & 4.7 & 653 & 23.5 & 459 & 681 & -- \\
\addlinespace[2pt]
No-execution & 31.4 & 0.1 & 3.5 & 491 & 26.4 & 422 & 521 & \textbf{0.68\xx} & 39.1 & $<$0.1 & 4.1 & 526 & 22.0 & 382 & 552 & \textbf{0.81\xx} \\
EET & 31.4 & 0.1 & 4.2 & 770 & 29.2 & 572 & 803 & 1.05\xx\ns & 37.9 & $<$0.1 & 4.5 & 619 & 23.1 & 459 & 646 & 0.95\xx\ns \\
SHERLOC & 32.5 & 0.1 & 7.9 & 902 & 108.8 & 884 & 1,018 & 1.33\xx & 41.4 & $<$0.1 & 6.0 & 523 & 57.6 & 380 & 587 & 0.86\xx\ns \\
AgentDiet & 29.0 & 46.6 & 30.8 & 651 & 65.8 & 645 & 794 & 1.04\xx\ns & 37.3 & 18.3 & 5.7 & 548 & 37.1 & 453 & 609 & 0.89\xx\ns \\
SeeRepo & -- & -- & -- & -- & -- & -- & -- & -- & -- & -- & -- & -- & -- & -- & -- & -- \\
\midrule
\flow CFG & 30.8 & 0.1 & 8.8 & 466 & 47.9 & 414 & 523 & \textbf{0.68\xx} & 42.0 & $<$0.1 & 6.0 & 606 & 29.9 & 503 & 642 & 0.94\xx\ns \\
\flow DFG & 29.6 & 0.1 & 9.2 & 504 & 51.1 & 457 & 565 & 0.74\xx & 39.6 & $<$0.1 & 5.7 & 594 & 29.3 & 492 & 629 & 0.92\xx\ns \\
\midrule
\multicolumn{17}{@{}l}{\emph{SWE-bench Verified ($n=218$)}} \\
Baseline & 61.0 & 0.3 & 12.0 & 3,351 & 50.9 & 3,381 & 3,414 & -- & 73.9 & $<$0.1 & 3.5 & 781 & 18.5 & 578 & 803 & -- \\
\addlinespace[2pt]
No-execution & 57.3 & 0.2 & 9.3 & 2,384 & 41.0 & 2,395 & 2,434 & 0.71\xx & 75.2 & $<$0.1 & 2.3 & 429 & 15.5 & 264 & 447 & \textbf{0.56\xx} \\
EET & 62.8 & 0.3 & 12.2 & 2,998 & 66.2 & 2,823 & 3,076 & 0.90\xx & 77.1 & $<$0.1 & 3.8 & 856 & 19.8 & 600 & 879 & 1.10\xx\ns \\
SHERLOC & 60.1 & 0.3 & 13.1 & 2,756 & 117.7 & 2,773 & 2,887 & 0.85\xx & 73.9 & $<$0.1 & 4.1 & 665 & 43.3 & 340 & 713 & 0.89\xx\ns \\
AgentDiet & 58.3 & 94.5 & 72.3 & 2,416 & 128.0 & 2,579 & 2,710 & 0.79\xx & 76.6 & 9.6 & 4.5 & 708 & 26.1 & 521 & 749 & 0.93\xx\ns \\
SeeRepo & 61.9 & 0.3 & 13.7 & 3,782 & 59.5 & 4,102 & 3,856 & 1.13\xx & 76.1 & $<$0.1 & 3.7 & 840 & 25.1 & 553 & 869 & 1.08\xx\ns \\
\midrule
\flow CFG & 58.7 & 0.2 & 9.4 & 1,021 & 60.7 & 710 & 1,091 & \textbf{0.32\xx} & 70.6 & $<$0.1 & 4.2 & 560 & 20.7 & 407 & 585 & 0.73\xx \\
\flow DFG & 56.4 & 0.2 & 10.2 & 1,108 & 67.2 & 765 & 1,185 & 0.35\xx & 73.4 & $<$0.1 & 4.2 & 573 & 20.8 & 379 & 598 & 0.74\xx \\
\bottomrule
\end{tabular}
\end{table*}

\begin{figure*}[t]
\centering
\includegraphics[width=\textwidth]{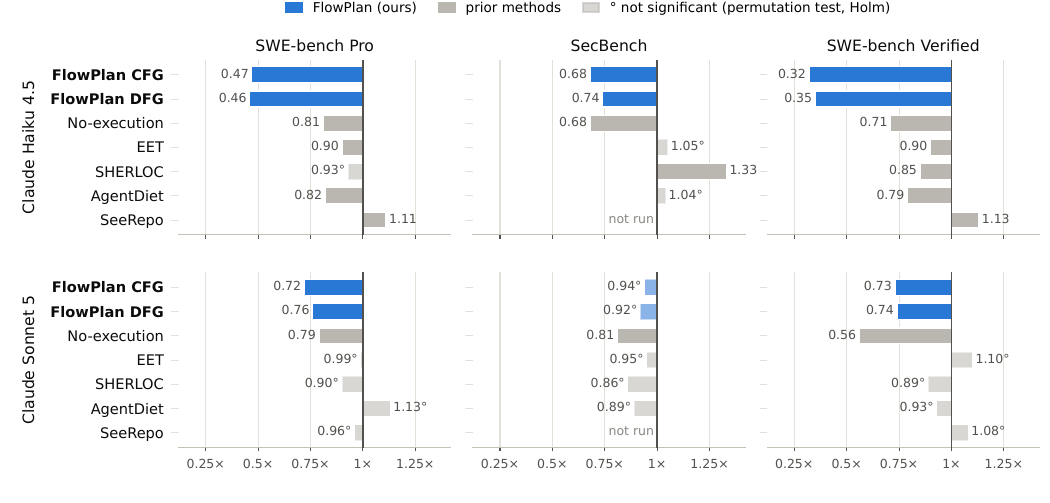}
\caption{Token consumption of each method relative to the vanilla agent, by benchmark (columns) and model (rows). Tokens include input, output, cache-read, and cache-write tokens. The baseline is $1\times$, and values below $1\times$ (bars to the left) indicate fewer tokens. Faded bars and values marked \ns{} do not differ significantly from the baseline under a paired permutation test of the mean difference with Holm correction. SeeRepo supports only Python and is therefore not evaluated on SecBench.}
\label{fig:tokenratios}
\end{figure*}

Table~\ref{tab:main} reports task success and token consumption, broken down into input, output, cache-read, and cache-write tokens, for all methods, while Figure~\ref{fig:tokenratios} compares their token ratios directly. 

\subsection{RQ1: Effectiveness and Cost}
\label{sec:results:rq1}

\textit{\textbf{\flow preserves task success.}}
Across the twelve \flow configurations (two graph views across six benchmark--model settings), resolution differences relative to the vanilla agent range from $-4.6$ to $+4.5$ percentage points, with no consistent direction. For example, with \haiku, CFG increases resolution from 43.0\% to 46.8\% on SWE-bench Pro and from 29.0\% to 30.8\% on SecBench, but decreases it from 61.0\% to 58.7\% on SWE-bench Verified. DFG shows the same mixed pattern. None of the CFG or DFG differences are statistically significant under McNemar's exact test. Thus, within the power of our evaluation, the token reductions reported below are not accompanied by evidence of a systematic gain or loss in repair success.

\noindent\textit{\textbf{\flow reduces token consumption in every setting, with the largest reductions on longer trajectories.}} Both graph views use fewer tokens than the baseline in all twelve comparisons. With \haiku, \flow uses only 0.46--0.47\xx{} of the baseline tokens on SWE-bench Pro and 0.32--0.35\xx{} on SWE-bench Verified. With Sonnet~5, it uses 0.72--0.76\xx{} on SWE-bench Pro and 0.73--0.74\xx{} on SWE-bench Verified. SecBench shows smaller reductions: 0.68--0.74\xx{} with \haiku and 0.92--0.94\xx{} with Sonnet~5. Ten of the twelve reductions are statistically significant under the paired permutation test after Holm correction ($p<0.001$). The only non-significant comparisons are CFG and DFG with Sonnet~5 on SecBench. The pattern is consistent with the amount of trajectory available to eliminate: the largest reductions occur on the long \haiku SWE-bench sessions, whereas the smallest occur on the short Sonnet~5 SecBench sessions. CFG and DFG exhibit similar trends across the six settings, with neither abstraction consistently outperforming the other in resolution or token consumption.

\noindent\textit{\textbf{Reduced cache reads drive the token savings.}} Cache reads account for 96--98\% of the vanilla agent's tokens across all settings because each model call replays the accumulated conversation from the prompt cache. Accordingly, most of \flow's token reduction comes from fewer cache-read tokens. With \haiku, cache reads fall to 0.44--0.46\xx{} of the vanilla agent on SWE-bench Pro, 0.30--0.33\xx{} on SWE-bench Verified, and 0.64--0.69\xx{} on SecBench. In contrast, cache-write tokens increase because the \tl{/flowfix} skill, MCP tool descriptions, and returned graph views add new context to the session. These additional tokens offset only a small fraction of the cache-read savings on SWE-bench Pro and Verified (at most 1.3\% and 1.5\%, respectively), but a larger fraction on SecBench (9--16\%). On SecBench, output tokens also increase because \flow requires the agent to explicitly state its repair plan and graph deltas. Because SecBench trajectories are shorter, this additional overhead offsets a larger portion of the cache-read savings, resulting in smaller overall token reductions.

\begin{findingbox}
\textbf{RQ1.}
\flow reduces token consumption across all benchmark--model settings without a statistically significant change in resolution. The reductions are significant in all but the Sonnet~5 SecBench setting, where shorter trajectories leave less room for savings. Most of the savings come from reduced cache reads.
\end{findingbox}

\subsection{RQ2: Comparison with Prior Token-Reduction Methods}
\label{sec:results:rq2}

\noindent\textit{\textbf{\flow provides the strongest token reductions in most settings.}} Figure~\ref{fig:tokenratios} places methods on one scale. With \haiku, \flow uses 0.46--0.47\xx{} of the baseline tokens on SWE-bench Pro, compared with 0.81\xx{} for the strongest reproduced alternative, and 0.32--0.35\xx{} on SWE-bench Verified, compared with 0.71\xx. On SecBench, CFG and No-execution use 0.68\xx{} of the baseline tokens. With Sonnet~5, CFG remains lowest on SWE-bench Pro at 0.72\xx, while No-execution is lower on SWE-bench Verified and SecBench, where it's the only statistically significant reduction. Overall, \flow's advantage is most pronounced on the longer \haiku trajectories.

\noindent\textit{\textbf{No-execution is the strongest prior method, while extra model calls and context limit the savings of others.}}
No-execution significantly reduces token consumption in all six settings (0.56--0.81\xx) and uses fewer tokens than \flow in two settings, both with Sonnet~5. The remaining methods provide smaller or less consistent savings. SHERLOC's localization stage and AgentDiet's reflection and history-rewriting steps introduce model calls and cache writes, offsetting much of the context they remove; their total token use ranges from 0.85--1.33\xx{} and 0.79--1.13\xx{} of the baseline, respectively. EET remains close to the baseline (0.90--1.10\xx{}) and significantly reduces tokens only with \haiku on the two SWE-bench benchmarks, while SeeRepo uses 0.96--1.13\xx{} of the baseline. These results indicate that token reduction depends not only on shortening the repair context, but also on avoiding extra calls and cache updates that can cancel out those savings.

\noindent\textit{\textbf{The largest savings come from reducing repeated interactions rather than adding preprocessing.}} The two methods with the strongest overall token reductions, \flow and No-execution, both reduce substantial portions of the agent's interaction trajectory, but in different ways. No-execution removes repeated program execution, whereas \flow uses a focused program abstraction for pre-edit reasoning and returns to the same abstraction after editing to verify the intended change. By contrast, SHERLOC, AgentDiet, and SeeRepo introduce additional localization, compression, or representation machinery whose overhead can offset the context they save.

\noindent\textit{\textbf{Resolution is largely unchanged across the compared methods.}} EET with Sonnet~5 on SWE-bench Pro is the only reproduced configuration whose resolution differs significantly from the baseline, increasing from 52.5\% to 63.3\% ($p=0.002$). No other method--benchmark--model combination shows a statistically significant resolution difference. The token comparisons can therefore generally be interpreted without a corresponding statistically detectable change in task success, with this EET result as the exception.

\begin{findingbox}
\textbf{RQ2.} \flow achieves the largest token reduction in three of the six benchmark--model settings and ties No-execution in a fourth. Its advantage is most pronounced on the longer \haiku trajectories. No-execution is the strongest reproduced alternative and outperforms \flow in the two remaining settings, both with Sonnet~5. Overall, methods that eliminate repeated agent interactions provide larger end-to-end savings than methods whose context reductions require substantial auxiliary computation.
\end{findingbox}

\subsection{RQ3: Analysis}
\label{sec:results:rq3}

\begin{table*}[t]
\centering
\caption{RQ3: Correctness of each step of the \flow protocol, judged against the reference patch (\haiku; SWE-bench Pro $n=158$, SWE-bench Verified $n=218$, SecBench $n=169$). Each cell gives the number of sessions and, in parentheses, their percentage of the benchmark sample. Marks: \checkmark{} correct, $\sim$ partly correct, $\times$ incorrect (deviates from the reference), $\circ$ not applicable because the fix lies outside what the pre-repair graph can show (new code, or no function changed), -- not applicable because the step was not taken. All: the sessions of each category pooled over the three benchmarks and both graph views ($n=1{,}090$).}
\label{tab:rq3steps}
\footnotesize
\setlength{\tabcolsep}{2.5pt}
\begin{tabular}{@{}cl rr rr rr l@{}}
\toprule
& & \multicolumn{2}{c}{SWE-bench Pro} & \multicolumn{2}{c}{SWE-bench Verified} & \multicolumn{2}{c}{SecBench} & \\
\cmidrule(lr){3-4}\cmidrule(lr){5-6}\cmidrule(lr){7-8}
& Outcome & CFG & DFG & CFG & DFG & CFG & DFG & All \\
\midrule

\multicolumn{9}{@{}l}{\textbf{Verify function localization}} \\
\checkmark & right function first & 109 (69\%) & 116 (73\%) & 120 (55\%) & 111 (51\%) & 89 (53\%) & 92 (54\%) & \multirow{2}{*}{\checkmark{} 842 (77\%)} \\
\checkmark & right function after search & 27 (17\%) & 26 (16\%) & 58 (27\%) & 63 (29\%) & 13 (8\%) & 18 (11\%) & \\
$\times$ & right file, different function & 7 (4\%) & 8 (5\%) & 11 (5\%) & 13 (6\%) & 24 (14\%) & 19 (11\%) & \multirow{2}{*}{$\times$ 192 (18\%)} \\
$\times$ & different function & 4 (3\%) & 0 (0\%) & 14 (6\%) & 16 (7\%) & 40 (24\%) & 36 (21\%) & \\
$\circ$ & reference only adds new code & 7 (4\%) & 7 (4\%) & 15 (7\%) & 15 (7\%) & 3 (2\%) & 3 (2\%) & \multirow{2}{*}{n/a 56 (5\%)} \\
-- & protocol not used & 4 (3\%) & 1 (1\%) & 0 (0\%) & 0 (0\%) & 0 (0\%) & 1 (1\%) & \\

\midrule
\multicolumn{9}{@{}l}{\textbf{Verify graph construction}} \\
\checkmark & faithful view & 108 (68\%) & 107 (68\%) & 193 (89\%) & 192 (88\%) & 143 (85\%) & 108 (64\%) & \multirow{2}{*}{\checkmark{} 913 (84\%)} \\
\checkmark & faithful; reference adds new code & 18 (11\%) & 22 (14\%) & 9 (4\%) & 13 (6\%) & 0 (0\%) & 0 (0\%) & \\
$\times$ & different function or empty view & 20 (13\%) & 22 (14\%) & 15 (7\%) & 13 (6\%) & 26 (15\%) & 60 (36\%) & $\times$ 156 (14\%) \\
$\circ$ & reference consists only of new code & 7 (4\%) & 6 (4\%) & 1 (0\%) & 0 (0\%) & 0 (0\%) & 0 (0\%) & \multirow{2}{*}{n/a 21 (2\%)} \\
-- & no graph requested & 5 (3\%) & 1 (1\%) & 0 (0\%) & 0 (0\%) & 0 (0\%) & 1 (1\%) & \\

\midrule
\multicolumn{9}{@{}l}{\textbf{Verify graph-element localization}} \\
\checkmark & reference-changed element & 80 (51\%) & 96 (61\%) & 124 (57\%) & 120 (55\%) & 42 (25\%) & 48 (28\%) & \checkmark{} 510 (47\%) \\
$\times$ & elsewhere in the graph & 31 (20\%) & 26 (16\%) & 68 (31\%) & 73 (33\%) & 124 (73\%) & 117 (69\%) & $\times$ 439 (40\%) \\
$\circ$ & reference inserts new code & 13 (8\%) & 13 (8\%) & 24 (11\%) & 24 (11\%) & 1 (1\%) & 1 (1\%) & \multirow{3}{*}{n/a 141 (13\%)} \\
-- & not pinned (optional) & 30 (19\%) & 22 (14\%) & 2 (1\%) & 1 (0\%) & 0 (0\%) & 0 (0\%) & \\
-- & protocol not used or no function & 4 (3\%) & 1 (1\%) & 0 (0\%) & 0 (0\%) & 2 (1\%) & 3 (2\%) & \\

\midrule
\multicolumn{9}{@{}l}{\textbf{Verify repair planning}} \\
\checkmark & matches reference & 57 (36\%) & 53 (34\%) & 108 (50\%) & 99 (45\%) & 52 (31\%) & 52 (31\%) & \checkmark{} 421 (39\%) \\
$\sim$ & right target, partial change & 66 (42\%) & 80 (51\%) & 21 (10\%) & 29 (13\%) & 16 (9\%) & 12 (7\%) & $\sim$ 224 (21\%) \\
$\times$ & different change & 25 (16\%) & 21 (13\%) & 88 (40\%) & 89 (41\%) & 97 (57\%) & 100 (59\%) & $\times$ 420 (38\%) \\
-- & no plan or no function & 10 (6\%) & 4 (3\%) & 1 (0\%) & 1 (0\%) & 4 (2\%) & 5 (3\%) & n/a 25 (2\%) \\

\midrule
\multicolumn{9}{@{}l}{\textbf{Verify plan realization}} \\
\checkmark & matches plan & 138 (87\%) & 140 (89\%) & 207 (95\%) & 203 (93\%) & 156 (92\%) & 145 (86\%) & \checkmark{} 989 (91\%) \\
$\times$ & does not match plan & 6 (4\%) & 7 (4\%) & 2 (1\%) & 6 (3\%) & 9 (5\%) & 17 (10\%) & $\times$ 47 (4\%) \\
-- & not checkable or no plan & 14 (9\%) & 11 (7\%) & 9 (4\%) & 9 (4\%) & 4 (2\%) & 7 (4\%) & n/a 54 (5\%) \\

\midrule
& \textbf{All steps agree with reference} & 30 (19\%) & 31 (20\%) & 80 (37\%) & 69 (32\%) & 20 (12\%) & 13 (8\%) & 243 (22\%) \\
& \quad resolved & 16 & 17 & 64 & 56 & 11 & 7 & 171 \\

\bottomrule
\end{tabular}
\end{table*}

\begin{figure}[t]
\begin{tcolorbox}[enhanced, colback=white, colframe=black, boxrule=0.5pt, arc=1pt,
                  left=3pt, right=3pt, top=1.5pt, bottom=1.5pt]
\scriptsize
\textbf{Issue} (\tl{django-13089}, SWE-bench Verified). The database cache backend sporadically fails with \emph{'NoneType' object is not subscriptable} in \tl{\_cull}.

\vspace{1pt}
\setlength{\aboverulesep}{0.3ex}\setlength{\belowrulesep}{0.4ex}
\begin{tabularx}{\linewidth}{@{}>{\raggedright\arraybackslash}p{0.17\linewidth}>{\raggedright\arraybackslash}X>{\raggedright\arraybackslash}p{0.17\linewidth}@{}}
\toprule
Step & Tool call and result & Category \\
\midrule
Function localization & \tl{flow\_of(\_cull)}: CFG of \tl{DatabaseCache.\_cull}, the function the fix changes & right function first \\
Graph construction & 6 blocks, 7 edges, faithful to the source (excerpt below) & faithful graph \\
Graph-element localization & \tl{flow\_locate(db.py:272)}: block B5, which subscripts the fetched row & on the buggy element \\
Repair planning & \tl{flow\_plan}: \tl{add\_guard} on \tl{DatabaseCache.\_cull} at line 272 & matches the reference \\
Plan realization & adds \tl{if row is not None}; \tl{flow\_check}: \tl{add\_guard} as planned, \textsc{exact} & matches the plan \\
Final verification & \tl{flow\_verify}: all deltas matched, no unplanned edit; the agent stops & resolved \\
\bottomrule
\end{tabularx}

\vspace{1pt}
\begin{tcolorbox}[colback=gray!6, colframe=gray!6, boxrule=0pt, arc=0pt, left=2pt, right=2pt, top=0.5pt, bottom=0.5pt]
\scriptsize\ttfamily
B2 L256 if self.\_cull\_frequency == 0: \quad T$\rightarrow$B3, F$\rightarrow$B4\\
B4 L259 connection = connections[db] \quad T$\rightarrow$B5, F$\rightarrow$exit\\
B5 L266 cull\_num = num // self.\_cull\_frequency \quad $\rightarrow$exit
\end{tcolorbox}

\textbf{Cost.} \flow: 10 model calls, 0.19M tokens. Baseline: 45 model calls, 2.32M tokens (also resolved).
\end{tcolorbox}
\vspace{-6pt}
\caption{A session in which every protocol step is correct (\flow CFG, \haiku). Each step is judged against the reference patch, which adds the same guard.}
\label{fig:rq3example}
\end{figure}

RQ1 and RQ2 establish that \flow can substantially reduce token consumption without a statistically significant change in repair success. RQ3 asks why. We analyze all \haiku trajectories for both CFG and DFG views from two complementary perspectives: 1) whether each stage of the protocol performs its intended role relative to the reference patch, and 2) where tokens are consumed along the trajectory and which phases account for the savings.

\noindent\textit{\textbf{Does the agent follow the protocol correctly?}}
For every session, we reconstruct the protocol decisions from the tool calls and compare each with the reference patch. Table~\ref{tab:rq3steps} decomposes each trajectory into five observable decisions: whether \tl{flow\_of} reaches a function modified by the reference patch (\emph{function localization}); whether the returned graph faithfully represents that function (\emph{graph construction}); whether \tl{flow\_locate} identifies the graph element changed by the reference (\emph{graph-element localization}); whether the planned graph deltas correspond to those induced by the reference patch (\emph{repair planning}); and whether the subsequent edit matches the agent's own plan (\emph{plan realization}). For each decision, we classify a session as correct, incorrect, or not applicable. A decision is not applicable when the pre-repair graph cannot represent the reference change, such as newly introduced code, or when the agent omits an optional step; for repair planning, we additionally distinguish partly correct plans. The last column of the table pools the sessions of all three benchmarks and both graph views. Note that disagreement with the reference patch does not necessarily imply an incorrect repair: a semantically valid patch may implement the fix in a different function or through a different graph transformation. We therefore use these categories to analyze protocol behavior rather than as an alternative correctness oracle.

\noindent\textit{\textbf{Function localization and graph construction usually succeed, but SecBench is harder.}} Pooled over all sessions, function localization is correct in 77\% of sessions and graph construction in 84\%. On SWE-bench Pro, the agent opens a function changed by the reference patch in 86--90\% of sessions, and the returned graph faithfully represents the relevant function in 80--82\%; on SWE-bench Verified, these rates are 80--82\% and 93--94\%. On SecBench, these rates fall to 60--65\% for localization and 64--85\% for graph fidelity. A major reason is the crash stack: the sanitizer identifies where the failure manifests, but the reference repair may modify a caller or another function responsible for validating the input. The agent therefore often begins from a plausible crash location even when the eventual repair lies elsewhere.

\noindent\textit{\textbf{Graph-element localization is the weakest localization step.}} Pooled over all sessions, the agent pins the graph element changed by the reference patch in only 47\% of sessions. On SWE-bench Pro and Verified, \tl{flow\_locate} identifies this element in 51--61\% and 55--57\% of all sessions, respectively, although this rises to 72--79\% and 62--65\% when the agent pins a site on a graph that actually contains the reference change. Some remaining cases cannot be pinned because the repair introduces new code rather than modifying an existing graph element. SecBench is more difficult: the agent frequently pins the crash statement reported by the sanitizer even when the repair belongs elsewhere in the function or call stack. This gap illustrates an important distinction between locating a failure symptom and locating the code that must change to remove its cause.

\noindent\textit{\textbf{Repair planning is also difficult, but once a plan is stated, the agent usually realizes it.}} On SWE-bench Pro, 78--84\% of plans either match the graph change induced by the reference patch or capture part of that change in the correct function. On SWE-bench Verified and SecBench, however, 40--59\% of plans describe a different graph change. Some differences reflect genuine planning errors, while others are alternative repairs: for example, 19 of the 97 SecBench CFG sessions whose plan differs from the reference still resolve the vulnerability. In contrast, plan realization is highly reliable: once a plan exists, \tl{flow\_check} confirms that the edit matches it in 86--95\% of sessions across all three benchmarks. Pooled over all sessions, 59\% of plans match the reference or are partly correct, whereas 91\% of edits realize the agent's plan. The main difficulty is therefore formulating the correct repair at the abstract level. Once that intent has been made explicit, the agent usually succeeds in carrying it out in the concrete source code.

\noindent\textit{\textbf{End-to-end protocol correctness is associated with substantially higher repair success.}} Every step agrees with the reference in 19--20\% of SWE-bench Pro sessions, 32--37\% of SWE-bench Verified sessions, and 8--12\% of SecBench sessions, depending on graph view. These end-to-end correct trajectories resolve 53--81\% of their instances, compared with 28--46\% among the remaining sessions. This shows that the protocol's intermediate decisions are meaningful indicators of whether the repair is on track. Figure~\ref{fig:rq3example} illustrates one such trajectory: \flow identifies the relevant function, pins the faulty block, plans the guard introduced by the reference patch, verifies that the edit produces that change, and terminates after 10 calls and 0.19M tokens, compared with 45 calls and 2.32M tokens for the also-successful baseline.

\begin{findingbox}
\textbf{RQ3 (protocol correctness).} The agent usually reaches a function changed by the reference patch (77\% of sessions) and obtains a faithful graph of it (84\%), and once it records a repair plan, the subsequent edit realizes that plan in 91\% of sessions. The more difficult steps are identifying the exact graph element to change and formulating a plan that matches the reference repair.
\end{findingbox}

\begin{figure*}[t]
\centering
\includegraphics[width=\textwidth]{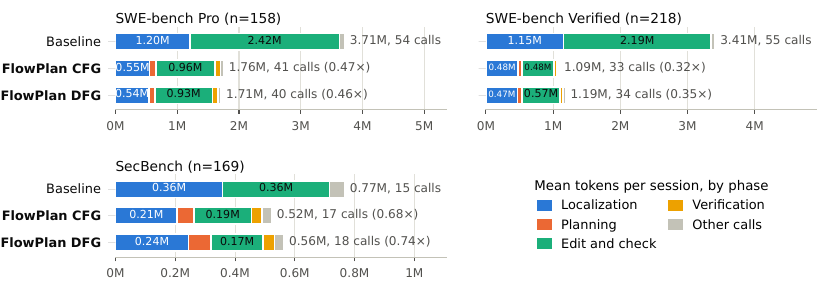}
\caption{RQ3: where tokens are consumed (\haiku, both \flow views). Mean tokens per session are decomposed by trajectory phase. Baseline sessions are split at the first edit. \flow calls are assigned to their protocol phase. Labels report total tokens, mean model calls, and the ratio to the baseline.}
\label{fig:rq3phases}
\end{figure*}

\noindent\textit{\textbf{Where are the tokens spent and saved?}}
The token cost of a session is the number of model calls times the context each call replays. To attribute it, we split every baseline session at its first edit into a localization phase and an edit-and-check phase, and assign every \flow call to the protocol phase of its tool call (Figure~\ref{fig:rq3phases}).

\noindent\textit{\textbf{Most SWE-bench savings arise after the first edit.}} 
In the vanilla agent, roughly two-thirds of the total tokens on SWE-bench Pro and Verified are consumed after the first substantive edit: 2.42M and 2.19M tokens per session, respectively. This post-edit tail contains about 31 additional model calls in each benchmark, during which the agent repeatedly re-reads source files, inspects the diff, executes tests, and performs further checks before deciding to stop. Because these interactions occur late in the trajectory, they are particularly expensive: each call replays accumulated context. With CFG, \flow reduces this phase to 18 and 12 calls and to 0.96M and 0.48M tokens, respectively. Consequently, the post-edit phase alone accounts for 75\% of the total token reduction on Pro and 74\% on Verified. DFG exhibits the same pattern, consuming 0.93M and 0.57M tokens after the first edit, with 75\% and 73\% of its savings originating in this phase. The reduction therefore comes not merely from making individual observations smaller, but from shortening an otherwise long and increasingly expensive verification tail: once \tl{flow\_check} establishes whether an edit produces the planned delta and \tl{flow\_verify} confirms the complete plan, the agent has an explicit completion criterion instead of continuing with repeated source inspection, diff checking, and test execution.

\noindent\textit{\textbf{\flow also reduces the exploratory prefix, while planning and verification add comparatively little overhead.}} With CFG, pre-edit token consumption falls from 1.20M to 0.55M on SWE-bench Pro and from 1.15M to 0.48M on Verified. The agent reaches the relevant code in fewer calls, and per-function graph observations introduce less accumulated context than repeated source-file and shell observations. The additional planning and verification phases require only 2.8--3.7 model calls and 0.08--0.17M tokens per session, less than one tenth of the tokens they ultimately save. Thus, the largest reduction does not come from making the new protocol itself cheaper, but from the larger exploratory and post-edit loops that the protocol prevents.

\noindent\textit{\textbf{SecBench explains when the mechanism has less room to help.}} The baseline SecBench trajectory is only about 15 calls long and divides its token use relatively evenly before and after the first edit. \flow therefore cannot remove a long post-edit tail: planning and verification instead increase the trajectory slightly to about 17--18 calls. Its token reduction comes primarily from smaller contexts per call(about 31K tokens versus 51K for the baseline) which reduces localization and post-edit consumption to 49--68\% of their baseline levels. This explains why \flow still lowers mean tokens on SecBench but achieves much smaller gains than on SWE-bench, particularly with Sonnet~5, where the difference is not statistically significant.

\begin{findingbox}
\textbf{RQ3 (token savings).} Most savings come from shortening the trajectory. After the first edit, \tl{flow\_check} and \tl{flow\_verify} replace much of the baseline's re-reading, diff inspection, and testing, accounting for 75\% of the reduction on SWE-bench Pro and 73--74\% on SWE-bench Verified. \flow also reduces pre-edit exploration, while its planning and verification cost is small relative to the savings. On SecBench, shorter trajectories leave less post-edit work to remove, so savings mainly come from reducing the context replayed by each model call.
\end{findingbox}
\section{Threats to Validity}
\label{sec:threats}

Each method is run once per instance, so outcomes may be affected by model nondeterminism. We mitigate this by evaluating methods on the same instances and using paired tests (McNemar's test for resolution and a permutation test of the mean difference for token consumption). Repeated runs would better estimate variability but would substantially increase the cost of evaluating multiple methods, models, and benchmarks. 
Our comparison with prior methods depends on reproduction fidelity. We retain released implementations, prompts, thresholds, and method-specific behavior where possible, but some approaches require integration or language-specific adaptation to run in a common \cc environment. Their absolute performance may therefore differ from the original studies, although evaluating all methods on the same instances, models, agent build, and accounting scheme supports consistent within-study comparisons. 
\flow uses CFGs and DFGs as repair abstractions, which may omit information needed for some fixes, such as cross-function context, semantics not captured by control or data flow, or newly introduced code. We therefore retain the agent's ordinary source-reading, search, and execution tools when the graph is insufficient. 
Finally, our results are limited to \cc, two models from the same provider, and three benchmarks covering Python repair and C/C++ security patching. Other agents, models, reasoning settings, languages, or tasks may exhibit different trajectories and savings. Our sampling targets a 95\% confidence level and 5\% margin of error within each benchmark, but does not establish generalization beyond those benchmarks.
\section{Conclusion}
\label{sec:conclusion}

We presented \flow, an abstraction-guided approach that structures coding-agent repair around per-function control- and data-flow graphs. \flow uses the graph as a common abstraction for fault localization, repair planning, and post-edit verification: the agent anchors the candidate fault location to the graph, expresses its intended repair as typed graph deltas, and verifies that the resulting program structure matches those deltas before terminating. Across \cc, two models, and three benchmarks, we find no statistically significant change in repair success relative to the vanilla agent, while \flow reduces token consumption in every benchmark--model setting and significantly in five of six. With \haiku, it uses 0.46--0.47\xx{} of the baseline tokens on SWE-bench Pro and 0.32--0.35\xx{} on SWE-bench Verified; with Sonnet~5, it uses 0.72--0.76\xx{} and 0.73--0.74\xx, respectively. It achieves the largest token reduction among the five reproduced 2026 methods in three of six settings and ties the strongest of them in a fourth. The trajectory analysis explains where these savings arise: \flow reduces pre-edit exploration, and after the first edit, where explicit plan verification replaces much of the baseline's repeated re-reading, diff inspection, and testing, it achieves 75\% of its token reduction on SWE-bench Pro and 73--74\% on SWE-bench Verified. The benefit is smaller on short SecBench trajectories, where there is less interaction to eliminate and the cache writes that \flow adds offset a larger share of the savings. More broadly, our results suggest that reducing coding-agent cost requires controlling not only what information enters the context, but also how the repair trajectory progresses and when the agent has sufficient evidence to stop.

\section*{Data Availability}
A replication package is available at 
\url{https://github.com/MelikaSepidband/FlowPlan}. It contains the \flow tools and the \tl{/flowfix} skill, the evaluation harness, our reproductions of the five compared methods, and the benchmark samples together with the sampling seed.

\bibliographystyle{ACM-Reference-Format}
\bibliography{ref}

\end{document}